\documentclass[aps,superscriptaddress,twocolumn,showpacs, showkeys,preprintnumbers,amsmath,amssymb,prb]{revtex4-1}
\usepackage{graphicx}
\usepackage{dcolumn}
\usepackage{bm}
\usepackage[dvips]{color}
\usepackage{natbib}
\usepackage{refstyle}

\begin{document}
\title{Anisotropic putative ``up-up-down'' magnetic structure in EuTAl$_4$Si$_2$ (T~=~Rh and Ir)}
\author{Arvind Maurya}
\affiliation{Department of Condensed Matter Physics and Materials
Science, Tata Institute of Fundamental Research, Homi Bhabha Road,
Colaba, Mumbai 400 005, India}
\author{P. Bonville}
\affiliation{CEA, Centre d'Etudes de Saclay, DSM/IRAMIS/Service de Physique de I'Etat Condens\'e, 91191 Gif-sur-Yvette, France}
\author{A. Thamizhavel}
\affiliation{Department of Condensed Matter Physics and Materials
Science, Tata Institute of Fundamental Research, Homi Bhabha Road,
Colaba, Mumbai 400 005, India}
\author{S. K. Dhar}
\email{sudesh@tifr.res.in}
\affiliation{Department of Condensed Matter Physics and Materials
Science, Tata Institute of Fundamental Research, Homi Bhabha Road,
Colaba, Mumbai 400 005, India}
\date{\today}

\begin{abstract}
We present detailed investigations in single crystals of two recently reported quaternary intermetallic compounds EuRhAl$_4$Si$_2$ and EuIrAl$_4$Si$_2$ employing magnetization, electrical resistivity in zero and applied fields, heat capacity and $^{151}$Eu M\"{o}ssbauer spectroscopy measurements. The two compounds order antiferromagnetically at $T_{\rm N1}$~=~11.7 and 14.7\,K, respectively, each undergoing two magnetic transitions: the first from paramagnetic to incommensurate modulated antiferromagnetic, the second at lower temperature to a commensurate antiferromagnetic phase as confirmed by heat capacity and M\"{o}ssbauer spectra. The magnetic properties in the ordered state present a large anisotropy despite Eu$^{2+}$ being an $S$-state ion for which the single-ion anisotropy is expected to be weak. Two features in the magnetization measured along the $c$-axis are prominent. At 1.8\,K, a ferromagnetic-like jump occurs at very low field to a value one third of the saturation magnetization (1/3 M$_0$) followed by a wide plateau up to 2\,T for T~=~Rh and 4\,T for T~=~Ir. At this field value, a sharp hysteretic spin\--flop transition occurs to a fully saturated state (M$_0$). Surprisingly, the magnetization does not return to origin when the field is reduced to zero in the return cycle, as expected in an antiferromagnet. Instead, a remnant magnetization 1/3 M$_0$ is observed and the magnetic loop around the origin shows hysteresis. This suggests that the zero field magnetic structure has a ferromagnetic component, and we present a model with up to third neighbor exchange and dipolar interaction which reproduces the magnetization curves and hints to an ``up-up-down'' magnetic structure in zero field. 
\end{abstract}
\pacs{71.20.Dg, 75.30.Fv, 75.10.Dg, 75.50.Ee, 71.20.Lp, 76.80.+y}
\keywords{EuRhAl$_4$Si$_2$, EuIrAl$_4$Si$_2$, single crystal, magnetization steps, antiferromagnetism.}
\maketitle
\section{Introduction}
The synthesis of single crystals of new quaternary compounds EuTAl$_4$Si$_2$ (T~=~Rh and Ir), using the Al\--Si binary eutectic as flux, has recently been reported~\cite{maurya2014synthesis}. The two compounds were found to adopt an ordered derivative of the ternary KCu$_4$S$_3$\--type tetragonal, $tP8$, $P4/mmm$ structure. The Eu, Al, Si and T atoms occupy respectively the K(1$a$), Cu(4$i$), S2(2$h$) and S1(1$a$) Wyckoff sites, which leads to  quaternary and truly stoichiometric 1:1:4:2 compounds. The local symmetry at the Eu site is fourfold axial ($4/mmm$). Preliminary magnetization data~\cite{maurya2014synthesis}, taken with the applied field parallel to [001] direction, revealed a divalent state for Eu ions in both compounds and a magnetic transition at $\approx$~12 and $\approx$~15\,K in the Rh and Ir analogs, respectively. Electronic structure calculations using the local spin density approximation~\cite{maurya2014synthesis} are in conformity with the divalent state of Eu ions.

In the present work, we probe in detail the magnetic behavior of these two intermetallic compounds along the principal crystallographic directions of the tetragonal cell, using the techniques of magnetization, heat capacity and electrical transport. Additional information about the magnetic transitions is derived from $^{151}$Eu M\"{o}ssbauer spectroscopy. We find a rather unusual and highly anisotropic magnetic response, which points to a probable ferrimagnetic-like spin arrangement, scarcely observed in intermetallics with divalent Eu. We show that such a ground magnetic structure can be predicted using a mean field model with three exchange integrals. 
\section{Experimental}
The details of the single crystal growth and structure have been reported in Ref.\onlinecite{maurya2014synthesis}. In the present work single crystals of LaTAl$_4$Si$_2$ (T~=~Rh and Ir), isostructural to Eu analogs, were also grown following the same protocol as adopted for the Eu compounds. The flux-grown crystals were oriented by means of the Laue diffraction technique in back\--reflection mode and cut appropriately using a spark erosion cutting machine. The magnetization as a function of temperature and magnetic field was measured in Quantum Design SQUID and VSM magnetometers between 1.8 and 300\,K. Heat capacity and electrical resistivity  in zero and applied fields were measured in a Quantum Design PPMS. $^{151}$Eu  M\"{o}ssbauer absorption spectra were taken at selected temperatures using a commercial $^{151}$SmF$_3$~$\gamma$\--ray source, mounted on a constant acceleration electromagnetic drive, in a standard liquid He cryostat.
\section{Results}
\subsection{Magnetic susceptibility}
%
\begin{figure*}[!]
\includegraphics[width=0.85\textwidth]{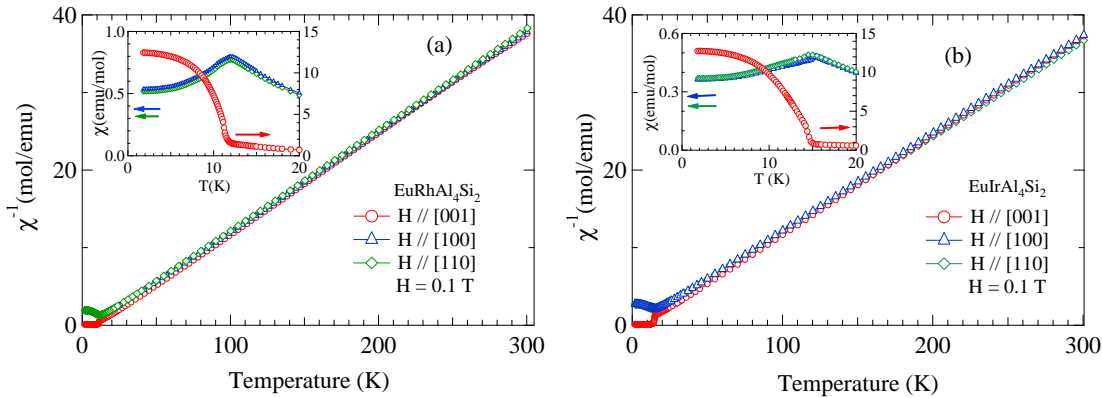}
\caption{\label{Chi_inverse_T}(Color online) Inverse magnetic susceptibility of (a) EuRhAl$_4$Si$_2$ and (b) EuIrAl$_4$Si$_2$ along [001], [100] and [110] crystallographic directions. The insets show the anisotropy in the low temperature susceptibility.}
\end{figure*}
\begin{figure*}[!]
\includegraphics[width=0.85\textwidth]{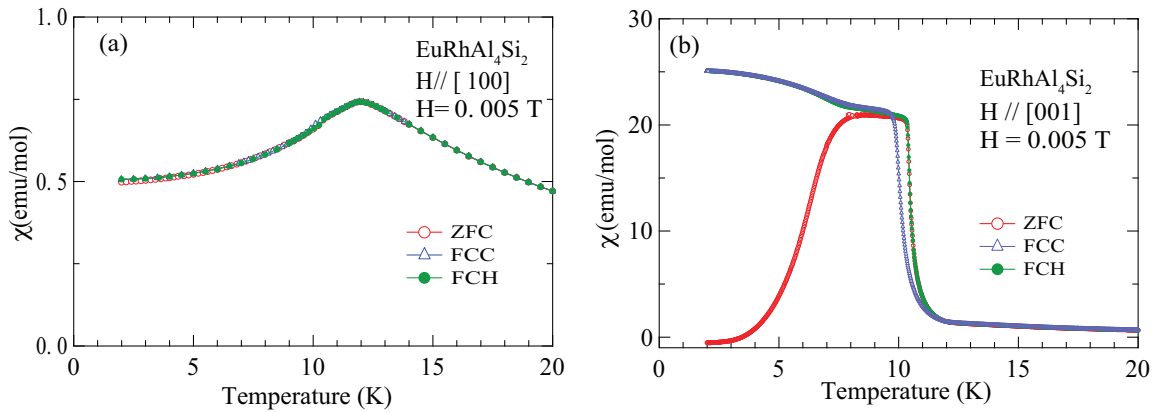}
\caption{\label{Chi_T_low_T}(Color online) The magnetic susceptibility of EuRhAl$_4$Si$_2$ in a field of 0.005\,T: zero field cooled (ZFC), field cooled cooling (FCC) and field cooled heating (FCH) datafor $H~\parallel$~[100] (a) and $H~\parallel$~[001] (b).}
\end{figure*}
Figures \ref{Chi_inverse_T}a and \ref{Chi_inverse_T}b show the inverse susceptibility data between 300 and 1.8\,K respectively for EuRhAl$_4$Si$_2$ and EuIrAl$_4$Si$_2$in a field of 0.1\,T applied  along the [001], [100] and [110] directions. The molar susceptibility follows the Curie-Weiss law:
\begin{equation}
\chi\left(T\right) = \frac{N\mu_{\rm eff}^2}{3k_B\left(T-\theta_p\right)},
\end{equation}
where $N$ is the Avogadro number, almost in the entire paramagnetic region. The effective moment $\mu_{\rm eff}$ and the Curie-Weiss temperature $\theta_p$ are listed in Table \ref{table1}. The $\mu_{\rm eff}$ values in both compounds are comparable to the Hund's rule derived value of 7.94\,$\mu_{\rm B}$ for free-ion Eu$^{2+}$. The positive $\theta_{\rm p}$ values in these {\it a priori} antiferromagnetic (AF) materials indicate the presence of competing nearest neighbor and next\--nearest exchange couplings, with opposite signs. This is probably brought about in these two metallic compounds by the oscillatory RKKY (Ruderman\--Kittel\--Kasuya\--Yosida) exchange interaction. An appreciable anisotropy in the susceptibility, plotted between 1.8 and 20\,K in the inserts of Fig.\ref{Chi_inverse_T}, develops at low temperatures in the two compounds.
\begin{table*}
\caption{\label{table1} Effective moment and paramagnetic Curie temperature values in EuRhAl$_4$Si$_2$ and EuIrAl$_4$Si$_2$ along the major crystallographic directions.}
\begin{ruledtabular}
\begin{tabular}{ccccc}
 &\multicolumn{2}{c}{EuRhAl$_4$Si$_2$} & \multicolumn{2}{c}{EuIrAl$_4$Si$_2$} \\ \hline
 & $\mu_{\rm eff}(\mu_{\rm B}/f.u.)$ & $\theta_p(K\textit{•})$ & $\mu_{\rm eff}(\mu_{\rm B}/f.u.)$ & $\theta_p(K\textit{•})$   \\ \hline
$H~\parallel$~[100] & $7.6$ & 8.0 & $7.8$ & 3.0 \\
$H~\parallel$~[110] & $7.84$ & $6.9$ & $8.1$ & $2.3$ \\
$H~\parallel$~[001] & $7.83$ & $10.4$ & $7.9$ & $9.2$ \\
\end{tabular}
\end{ruledtabular}
\end{table*}
%
It is clearly evidenced by the $\chi(T)$ data for EuRhAl$_4$Si$_2$ below 20\,K, taken in a low field of 0.005\,T, shown in Fig.\ref{Chi_T_low_T}a for $H~\parallel$~[100] and Fig.\ref{Chi_T_low_T}b for $H~\parallel$~[001]. 

The susceptibility $\chi_a(T)$ along the [100] direction (Fig.\ref{Chi_T_low_T}a) shows a peak at 12\,K followed by a minor kink at 10.4\,K (which correspond closely with the peak in zero field heat capacity at 11.7\,K and the small kink at 10.4\,K, {\it vide infra}) with nearly overlapping ZFC and FC data, characteristic of an AF transition.
In EuIrAl$_4$Si$_2$, the corresponding data (not shown) are similar, showing a peak at nearly 15\,K, followed by a barely discernible anomaly in the 13\--14\,K range (which also corresponds closely with the sharp peak in zero field heat capacity at 14.7\,K, followed by a peak at 13.2\,K, {\it vide infra}). Data taken with $H~\parallel$~[110] (not shown) reflect a nearly isotropic magnetic response in the $ab$-plane in both compounds. 
\begin{figure*}[!]
\includegraphics[width=0.85\textwidth]{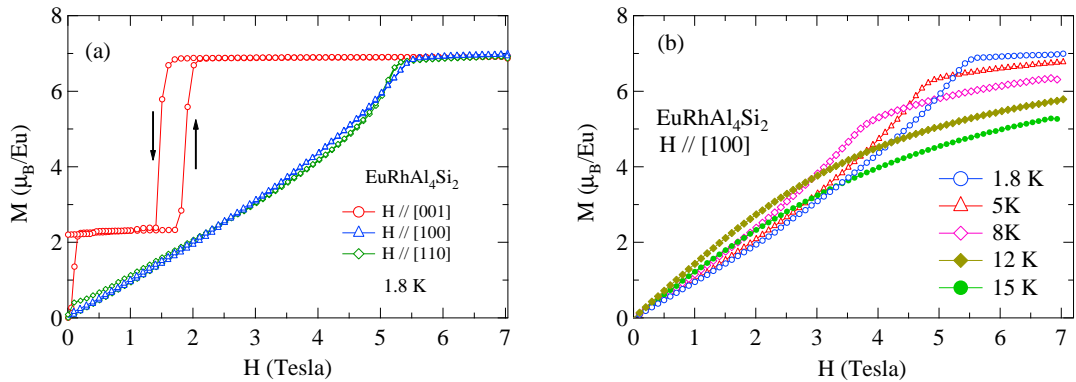}
\caption{\label{MH}(Color online) $M$ vs $H$ for EuRhAl$_4$Si$_2$ at 1.8\,K for $H$ in the $ab$-plane and along [001] (a) and at various temperatures for $H$ along [100] (b).}
\end{figure*}

On the other hand, along the tetragonal $c$-axis [001], the susceptibility $\chi_c(T)$ exhibits ferromagnetic\--like behavior, with its peak value more than an order of magnitude larger than the peak value of $\chi_a(T)$ (Fig.\ref{Chi_T_low_T}b). There is a substantial thermomagnetic irreversibility between ZFC and FC runs, which decreases as the applied field is increased (data at higher fields not shown here for the sake of clarity). Phenomenologically, the ZFC/FC behavior is similar to that seen in ferromagnets with a sizeable coercivity. The upturn in $\chi_c(T)$ begins at nearly the same temperature for $\chi_a(T)$. Data were taken in both field-cooled-cooling (FCC) and field-cooled-heating (FCH) modes. The observation of hysteresis in these two plots is in line with the ferromagnetic character of the transition for $H~\parallel$~[001]. Qualitatively, the data for EuIrAl$_4$Si$_2$ look similar and are not shown for brevity.

\subsection{Isothermal Magnetization}
\begin{figure*}[!]
\includegraphics[width=0.85\textwidth]{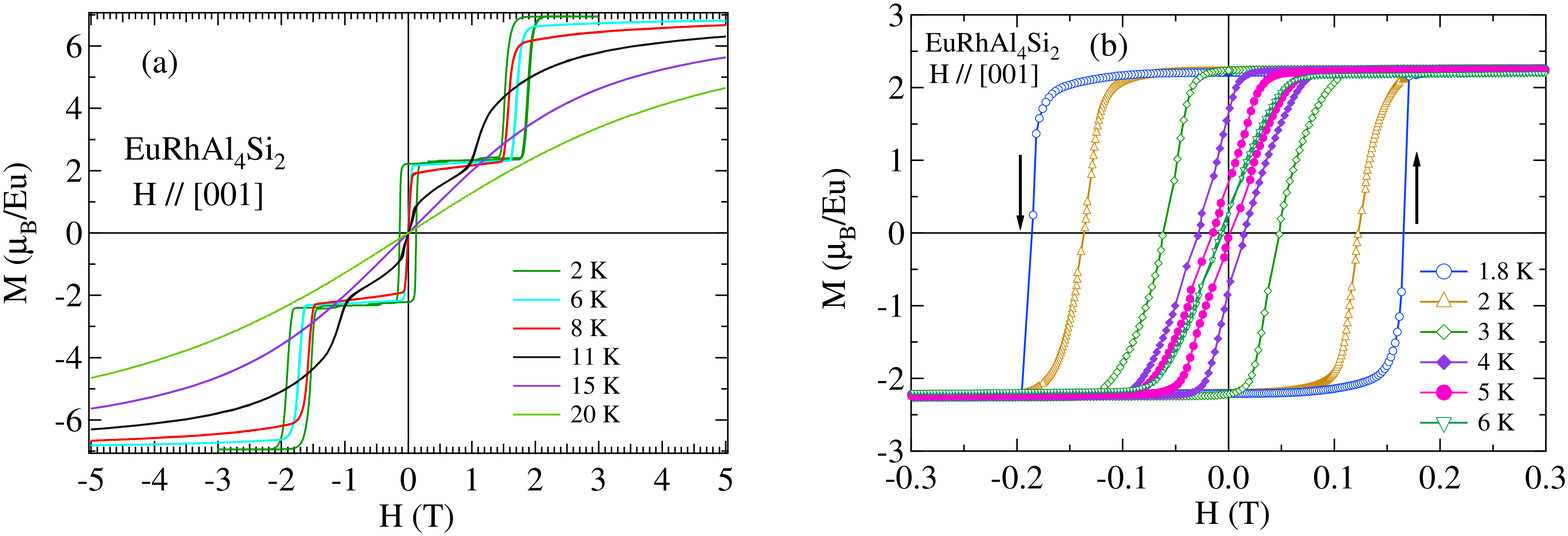}
\caption{\label{MH_hist}(Color online) $M$ {\it vs} $H$ curves in EuRhAl$_4$Si$_2$ at various temperatures showing the hysteretic behavior at low fields and at the spin-flop jump (a) and low field hysteresis on a reduced field scale showing the coercive field (b).}
\end{figure*}
The behavior of the in\--field magnetization in EuRhAl$_4$Si$_2$ at 1.8\,K, measured in the $ab$-plane along [100] and [110], and along [001] (Fig.\ref{MH}a), is in conformity with the $\chi(T)$ data depicted in Figs.\ref{Chi_T_low_T}a and \ref{Chi_T_low_T}b. For $H$ in the $ab$-plane, the magnetisation increases almost linearly, reaching a (spin\--flip) saturation field near 5.5\,T, with a saturated Eu$^{2+}$ moment M$_0$~=~7\,$\mu_{\rm B}$, as expected for a $g$=2, $S$=7/2 ion. This behavior is characteristic for an AF structure where the moments are perpendicular to the field.
At higher temperatures, both the spin-flip field value and the saturation magnetization decrease (see Fig.\ref{MH}b). By contrast, for $H~\parallel$~[001], there is a jump of the magnetization at very low field to a value 2.3\,$\mu_{\rm B}$, which is almost exactly 1/3 of M$_0$. The magnetization plateau is broken near 1.7\--2\,T by a hysteretic spin\--flop jump to saturation. This behavior tells that the Eu moment probably lies along the $c$-axis, and the low field jump hints to the presence of an average ferromagnetic component along $c$ in the zero field structure.
The 1/3 M$_0$ value suggests that the latter could consist in a motif of 3 spins along $c$, two pointing up and one pointing down. The zero field magnetic structure would then consist in ferromagnetic planes with moments along $c$ in an ``up-up-down'' sequence. The Ir sibling shows exactly the same behavior, but with a spin\--flop field of 3.9\--4\,T along [001] and a spin\--flip field of 9.2\,T perpendicular to [001].
%
\begin{table*}
\caption{\label{table2} Magnetic characteristics of EuRhAl$_4$Si$_2$ and EuIrAl$_4$Si$_2$.}
\begin{ruledtabular}
\begin{tabular}{ccccccc}
 & $T_{\rm N1}$ & $T_{\rm N2}$ & \multicolumn{2}{c}{Spin flip field at 1.8\,K (T)} & \multicolumn{2}{c}{Hysteresis at 1.8\,K $H~\parallel c$}\\ \hline
 & & & $H~\parallel ab$-plane & $H~\parallel c$ & $H_c (T)$ & Plateau width (T)\\ \hline 
EuRhAl$_4$Si$_2$ & 11.7 & 10.4 & 5.5 & 1.8 & 0.18 & 1.2\\
EuIrAl$_4$Si$_2$ & 14.7 & 13.2 & 9.2 & 3.9 & 0.25 & 2.5\\
\end{tabular}
\end{ruledtabular}
\end{table*}
%
Figure \ref{MH_hist}a shows the hysteretic behaviour of the magnetisation in EuRhAl$_4$Si$_2$, for both magnetisation jumps at $H$ close to origin and $H$ near 2\,T. Surprisingly, there is a sizeable remnant magnetization of 2.3\,$\mu_{\rm B}$/f.u. at 1.8\,K. 
Figure \ref{MH_hist}b shows the hysteresis loops at various temperatures in the narrow field interval of -0.3 to 0.3\,T. At 1.8\,K, there is a large coercive field of nearly 0.18\,T. The width of the loop depends sensitively on the temperature and it has decreased by more than half at 3\,K, compared to its value at 1.8\,K and by 6\,K the hysteresis has practically vanished. We suggest hysteresis is due to the presence of a ferromagnetic component in the zero field structure of these compounds, the possible domains having their mean magnetization pointing either up or down. The coercive field would then correspond to a complete reversal of the domains magnetization along the direction of the field. The primary sources of coercivity are the magnetocrystalline anisotropy and the dipolar field, which has been shown to play an important role in Eu compounds with low transition temperatures~\cite{Maurya2013EuNiGe3}. The rather high coercive field of 0.18\,T at 1.8\,K points thus to an unusually high crystalline anisotropy for an $L$=0 divalent Eu ion, for which it vanishes in principle at first order. This is supported by the modeling of the magnetic behavior presented in section \ref{disc}.
\begin{figure*}[!]
\includegraphics[width=0.85\textwidth]{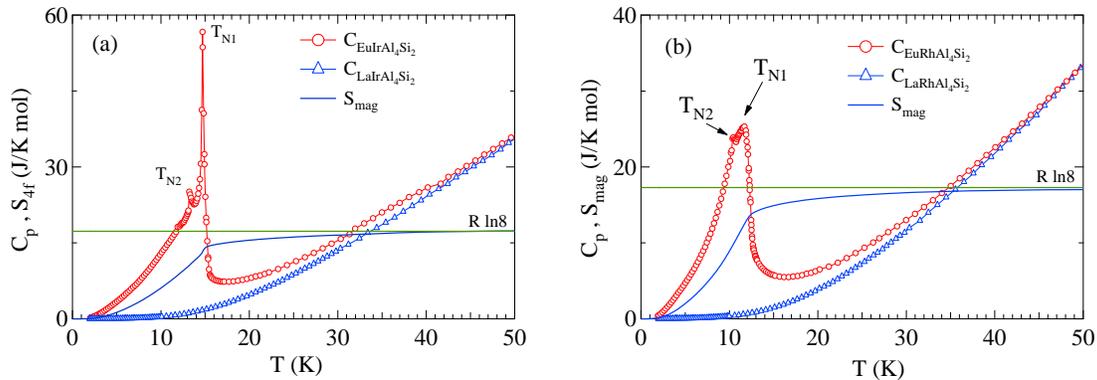}
\caption{\label{HC}(Color online) Heat capacity $C_p$ and entropy $S_{mag}$ as a function of temperature in EuTAl$_4$Si$_2$ and nonmagnetic LaTAl$_4$Si$_2$ for (a) T~=~Ir and (b) Rh. The horizontal line depicts the high temperature limit ($R\ln 8$) of the magnetic entropy for an $S=7/2$ spin system.}
\end{figure*}
\subsection{Heat Capacity}

The heat capacity of the two compounds is shown in Fig.\ref{HC}. We have also plotted the heat capacity of the corresponding La analogs, which serves as a measure of the phonon contribution assuming that the phonon spectra in the La and corresponding Eu compound are identical. This assumption appears valid in the present case as the heat capacities of La and Eu analogs almost coincide with each other at high temperatures (40\--50\,K) where the phonon contribution is dominant.

In EuIrAl$_4$Si$_2$, an extremely sharp peak with a peak value of nearly 60\,J/mol~K and peak position of 14.7\,K (Fig.\ref{HC}a), indicative of a first order phase transition from the paramagnetic state to an intermediate magnetically ordered state, is followed by a second transition near 13.2\,K marked by a small anomaly at that temperature. The heat capacity of EuRhAl$_4$Si$_2$ (Fig.\ref{HC}b) is apparently marked by only one peak at 11.7\,K which is relatively less sharp and has a lower peak height. However, a small kink at 10.4\,K is discernible; both are marked by arrows in the figure. The $^{151}$Eu M\"{o}ssbauer spectra, to be described below, also show the presence of two magnetic transitions in both compounds. The magnetic contribution to the heat capacity, $C_{mag}$, was obtained by subtracting the normalized heat capacity of the corresponding La compound, following the procedure described in Ref.\onlinecite{Blanco1991_spHeat}. 

The entropy $S_{mag}$ calculated by integrating $C_{mag}/T$ against the temperature is also shown in Figs.\ref{HC}a and \ref{HC}b. Nearly 75\% of the full entropy $R\ln 8$ is recovered at the transition temperature, and the remaining by $\simeq$~40\,K. The $C/T$ {\it vs} $T^2$ plots of LaRhAl$_4$Si$_2$ and LaIrAl$_4$Si$_2$ are linear below 5\,K. Fitting the expression $C/T=\gamma +\beta T^2$, where $\gamma$  and $\beta$ have their usual meaning, to the data furnishes $\gamma$~=~8.7 and 8\,mJ/molK$^2$ and $\beta$~=~0.2191 and 0.208\,mJ/molK$^4$ in the Rh and Ir compounds, respectively. The value of the Sommerfeld coefficient $\gamma$ ($\approx$ 1\,mJ/g atom) in these two compounds is comparable with those found in $sp$ and noble metals, and indicates a relatively  low density of states at the Fermi level compared to that typically found in $d$-transition metal alloys. The Debye temperature calculated from $\beta$ is 414 and 421\,K in LaRhAl$_4$Si$_2$ and LaIrAl$_4$Si$_2$, respectively. 
 
\subsection{Electrical Resistivity and Magnetoresistance}

\begin{figure*}[!]
\includegraphics[width=0.85\textwidth]{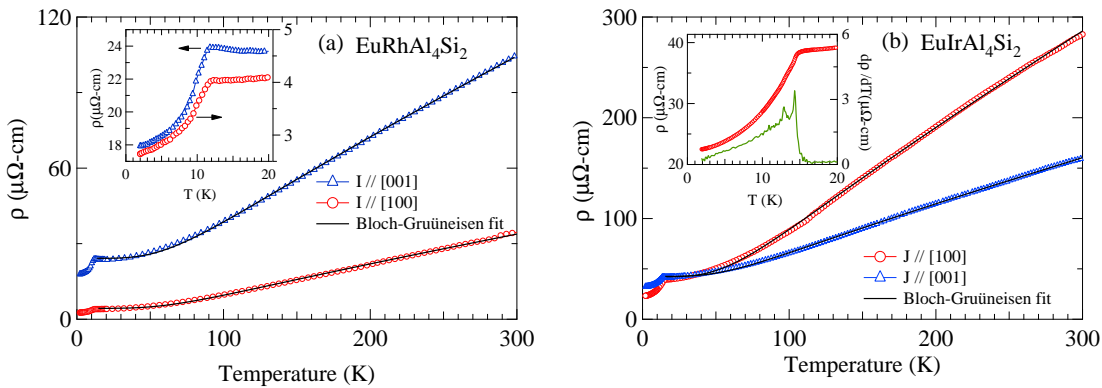}
\caption{\label{Resistivity_1}(Color online) Electrical resistivity as a function of temperature when the current is parallel to major crystallographic directions in (a) EuRhAl$_4$Si$_2$ and (b) EuIrAl$_4$Si$_2$.}
\end{figure*}

Figures \ref{Resistivity_1}a and \ref{Resistivity_1}b depict the variation of electrical resistivity $\rho(T)$ with temperature, which reveals normal metallic nature of the two compounds. The data have been taken with the current density along [100] and [001]. The resistivity is anisotropic which indicates an anisotropic Fermi surface of these two compounds with tetragonal symmetry. 
The fall in the resistivity at $T_{\rm N1}$, due to the gradual loss of spin disorder scattering below the magnetic transition is clearly visible. $T_{\rm N2}$ is marked by only a barely discernible kink in $\rho (T)$ plot but clearly observed in the derivative plot as shown for EuIrAl$_4$Si$_2$. The resistivity in the paramagnetic region was fitted to the following expression:
\begin{equation}
\rho(T) = \rho_0(T) +\rho_{sd} +B\left(T/\theta_R\right)^5\int^{\theta_R/T}_0\frac{x^5}{\left(e^x-1\right)\left(1-e^{-x}\right)}dx,
\end{equation}
where $\rho_0(T)$ is the residual resistivity, $\rho_{sd}$  is the resistivity contribution due to spin disorder, $B$ is a material specific parameter and $\theta_R$ is the Debye temperature. 
The third term is the Bloch-Gr\"uneisen expression for the resistivity due to electron\--phonon scattering. The solid lines drawn through the data points represent the fits derived from the above expression. The fit-coefficients are listed in Table 3. It may be noted that $\rho_{sd}$ in the mean field approximation is temperature independent in the paramagnetic region. To within the precision of our measurements we did not detect any hysteretic behavior of the resistivity when the temperature was cycled up and down across the magnetic transition.

\begin{figure*}
\includegraphics[width=0.85\textwidth]{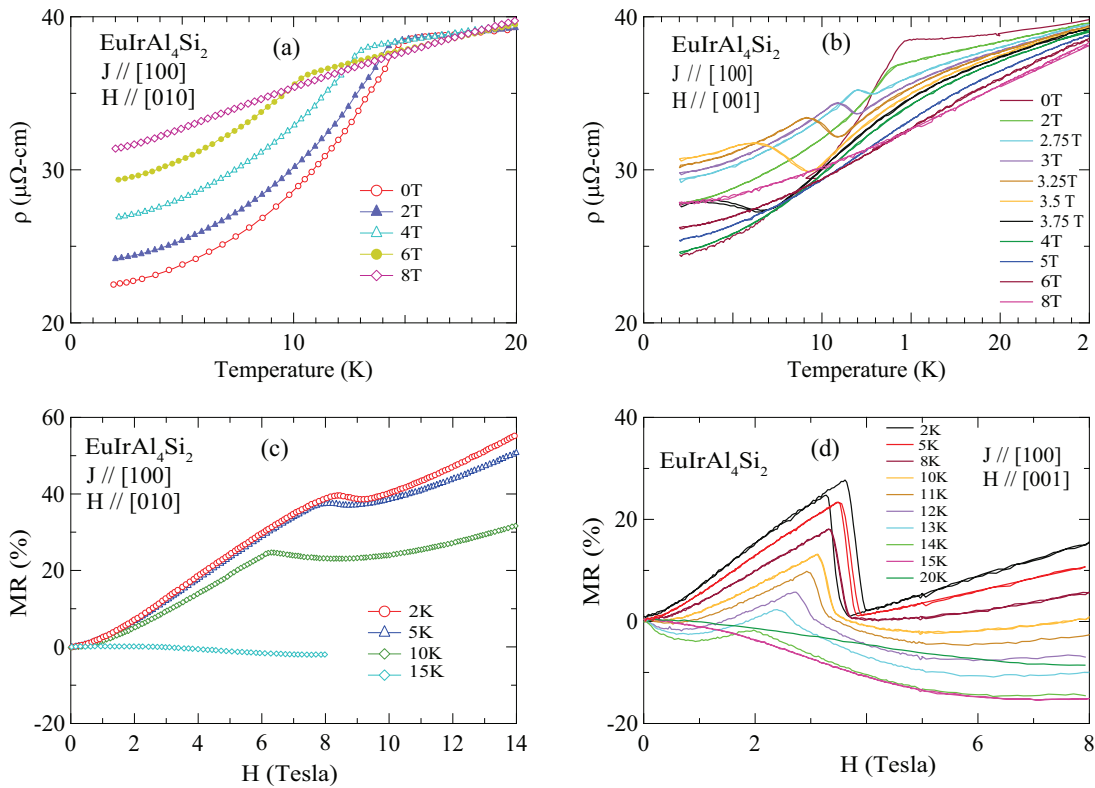}
\caption{\label{Resistivity_2}(Color online) $\rho (T)$ data as a function of temperature for different magnetic fields applied along (a) [010] and (b) [001], and field evolution of $MR$ for (c) $H~\parallel$~[010] and (d) $H~\parallel$~[001] in EuIrAl$_4$Si$_2$. The current is parallel to [100] in all the cases.}
\end{figure*}

The variation of resistivity with temperature in the magnetically ordered state in applied magnetic fields, and the magnetoresistance $MR$ derived from the data can be qualitatively understood on the basis of the in-field isothermal magnetization data presented in Figs. \ref{MH}a and \ref{MH}b. Figs.\ref{Resistivity_2}a and \ref{Resistivity_2}b show the resistivity below 25\,K in EuIrAl$_4$Si$_2$ with $J~\parallel$~[100] and $H~\parallel$~[010] and [001] at selected values of field between 0 and 8\,T. In Fig.\ref{Resistivity_2}a the main anomaly in the resistivity occurring at $T_{\rm N1}$ shifts towards lower temperatures with increasing field as expected for an antiferromagnet. The anomaly almost disappears at 8\,T as the spin\--flip field value is approached.
The Rh analog also shows similar behavior except that the anomaly at $T_{\rm N1}$ vanishes at a lower field consistent with a lower spin\--flip field in this compound when the field is applied in the $ab$-plane. In Fig.\ref{Resistivity_2}b, which shows the data for $H~\parallel$~[001], the main anomaly at $T_{\rm N1}$  again shifts to lower temperature with field. Additionally, the resistivity at 2.75, 3, 3.25, 3.5 and 3.75\,T shows an upturn at temperatures that is dependent on the applied field. The upturn, which suggests a field\--induced gap opening at the Fermi surface, corresponds nicely with the metamagnetic transitions observed in the magnetization when the field is applied along the $c$-axis. The magnetoresistance defined as $MR=\Delta R/R(0)$, where $\Delta R=R(H)-R(0)$, is shown in Figs.\ref{Resistivity_2}c and \ref{Resistivity_2}d for EuIrAl$_4$Si$_2$. The $MR$ of the Rh\--analogue is qualitatively similar (data not shown). The $MR$ of an antiferromagnetic material below $T_{\rm N}$ is expected to be positive as the applied field tends to break the antiferromagnetic couplings thereby increasing the spin disorder scattering.  Indeed, the $MR$ is positive below $T_{\rm N}$ in both the compounds and increases with increasing field. Near the field-induced spin-flip transition, the $MR$ tends to be field independent (Fig.\ref{Resistivity_2}c). However, at higher fields the positive $MR$ keeps on increasing though at a somewhat slower rate. With the increase of temperature the spin\--flip field decreases and this is reflected well in the $MR$ data of EuIrAl$_4$Si$_2$ at 2, 5 and 10\,K, respectively. Just above $T_{\rm N1}$ at 15\,K the $MR$ of EuIrAl$_4$Si$_2$ in the paramagnetic state is relatively small and negative in almost the entire range up to 8\,T. One may recall that the $MR$ due to the cyclotron motion of the free electrons is positive. Hence the negative, albeit small, $MR$ at 15\,K may be due to the partial quenching of the spin fluctuations above $T_{\rm N1}$. 

The $MR$ for $H~\parallel$~[001] and $J~\parallel$~[100] at selected values of temperature in fields up to 8\,T is shown in Figs.\ref{Resistivity_2}d. The prominent feature is the sharp drop and hysteresis in $MR$ at fields where the metamagnetic jump is seen in the magnetization of the Ir compound. The hysteresis is reduced in width and occurs at lower fields with the increase in temperature, in excellent correspondence with the magnetization data for $H~\parallel$~[001]. Thus there is a close correlation between the magnetization and the $MR$ data which clearly reveal the first order nature of the metamagnetic transition. Qualitatively the behavior of $MR$ in the two compounds is similar. At low fields, below $T_{\rm N}$, it is positive and increases with field. Above the metamagnetic transition the $MR$ increases with field, its magnitude and sign depending upon the temperature and field. In the paramagnetic state at 15\,K in EuRhAl$_4$Si$_2$ and at 20\,K in EuIrAl$_4$Si$_2$, the $MR$ is negative at all fields. 
%
\begin{table*}
\caption{\label{table3} Values of $\rho_0$, $\rho_{sd}$ and $\theta_R$ derived from the fit of equation 2 to the resistivity data in EuRhAl$_4$Si$_2$ and EuIrAl$_4$Si$_2$.}
\begin{ruledtabular}
\begin{tabular}{ccccc}
 & $\rho_0+\rho_{sd}$ &$\rho_0~at~2\,K$ & $\theta_R(K)$ & RRR \\
 & ($\mu\Omega cm$)  & ($\mu\Omega cm$)  & (K)  & $\rho(300\,K)/\rho(2\,K)$ \\\hline
 \multicolumn{5}{c}{EuRhAl$_4$Si$_2$} \\ \hline
$J~\parallel$~[100] &4.3 &2.6 & 384 & 12.7  \\
$J~\parallel$~[001] &24.2 &17.9 & 379 & 5.8  \\ \hline
 \multicolumn{5}{c}{EuRhAl$_4$Si$_2$} \\ \hline
$J~\parallel$~[100] &42.1 &22.6 & 354 & 12.5  \\
$J~\parallel$~[001] &42.5 &32.4 & 341 & 4.9  \\
\end{tabular}
\end{ruledtabular}
\end{table*}
%

\subsection{$^{151}$Eu M\"{o}ssbauer spectra}
\begin{figure}[h]
\includegraphics[width=0.45\textwidth]{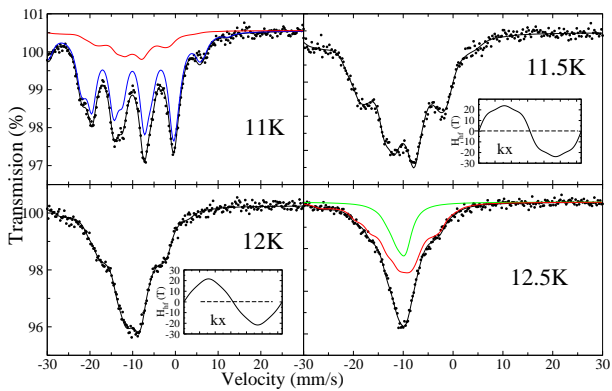}
\caption{\label{Mossbauer}(Color online) $^{151}$Eu M\"{o}ssbauer spectra in EuRhAl$_4$Si$_2$ between 11\,K and 12.5\,K (see text). The inserts at 11.5\,K and 12\,K show the modulation along the propagation vector \bf{k}.}
\end{figure}

Additional information on the two compounds was obtained by  employing $^{151}$Eu M\"{o}ssbauer spectroscopy. Spectra were recorded in EuRhAl$_4$Si$_2$ and in EuIrAl$_4$Si$_2$ between 4.2 and 14\,K. We show here only the data for T=Rh. We remind that the specific heat data show a transition near 11.7~K probably followed by another one at 10.4\,K. At 4.2\,K, the spectrum is a standard Eu$^{2+}$ hyperfine pattern with a single hyperfine field of 30.2(1)\,T, meaning that the magnetic structure is equal moment commensurate. 
The spectra start changing shape at 11\,K, and Figure \ref{Mossbauer} shows their rapid evolution in the small temperature interval 11-12.5\,K. At 11\,K, a superposition of a single hyperfine field spectrum (82\%, blue line) and of an incommensurate modulation (ICM) spectrum~\cite{Maurya2013EuNiGe3} (18\%, red line) is observed. At 11.5 and 12\,K, the spectra are purely ICM patterns, and at 12.5\,K the spectrum is a superposition of an ICM spectrum (72\%, red line) and of a single line paramagnetic spectrum (28\%, green line). At 14\,K, the spectrum is a single line characteristic of the paramagnetic phase.
Although the transitions we observe by M\"{o}ssbauer spectroscopy are slightly shifted (by 0.5\,K) with respect to those given by the specific heat peaks, the spectra illustrate well the way EuRhAl$_4$Si$_2$ goes, on cooling, from the paramagnetic phase to the ICM phase, then to the commensurate lock\--in phase, both transitions being first order as witnessed by the coexistence of the two phases in the spectra. The spectra in EuIrAl$_4$Si$_2$ are very similar, the transitions being shifted to somewhat higher temperature, in agreement with the specific heat data indicating a first transition close to 14.5\,K, followed by another at 13.2\,K. 
This cascade of transitions is a rather common feature in intermetallics with divalent Eu or trivalent Gd ions~\cite{Blanco1991_spHeat}. In particular, it is present in EuPdSb~\cite{bonville2001ICM_EuPdSb} and has been discovered more recently in EuPtSi$_3$~\cite{Neeraj2010EuPtSi3} and EuNiGe$_3$~\cite{Maurya2013EuNiGe3}. It seems to be concomitant with the occurence of a sizeable anisotropy, as in the present materials.
\section{Discussion} \label{disc}
Two noticeable features in the magnetization along the $c$-axis in both compounds are the existence of a jump at low field followed by a plateau region with magnetization 1/3 M$_0$ (M$_0$~=~7\,$\mu_{\rm B}$/f.u. at 2\,K). In a few rare earth antiferromagnets, plateaus in the magnetization along the easy-axis have been reported, but at a finite field value due to the strong crystal field induced Ising character of the considered rare earth. For example, in tetragonal DyCo$_2$Si$_2$ ($T_{\rm N}$~=~24\,K), the first spin\--flop occurs at 2\,T and the plateau magnetization is 5\,$\mu_{\rm B}$/Dy, exactly one half of the saturated value~\cite{iwata1990metamagnetism}. Neutron diffraction reveals that the intermediate magnetic structure consists of ferromagnetic (111) planes with the sequence +++- along the additional propagation vector Q2~=~(1/2, 1/2, 1/2)~\cite{shigeoka1994DyCo2Si2}. Orthorhombic DyCu$_2$ ($T_{\rm N}$~=~31\,K) exhibits a two\--step metamagnetic behavior with an intermediate magnetization of about 1/3 of the saturated value~\cite{iwata1989DyCu2Si2}. In CeZn$_2$ ($T_{\rm N}$~=~7\,K) there are two plateau regions with magnetizations of 1/3 and 1/2 of the saturation magnetization, starting respectively at 2\,T and 4\,T. Neutron diffraction on a single crystal of Ce(Zn$_0.9$Cu$_0.1$)$_2$, which is magnetically similar to CeZn$_2$, shows that the intermediate phases correspond to successive flipping of blocks of moments such that the magnetic periodicity is first tripled and then doubled along the easy $c$-axis~\cite{Gignoux_PD}. In hexagonal PrRh$_3$B$_2$, which shows the intermediate 1/3M$_0$ state, it was proposed that the ferromagnetic linear chains along [0001] are coupled antiferromagnetically, which leads to an Ising\--spin structure in a triangular lattice~\cite{yamada2004PrRh3B2}. 

To our knowledge, the only very low field 1/3 M$_0$ plateau has been observed at very low temperature for a field along [111] in spin\--ice Dy$_2$Ti$_2$O$_7$~\cite{matsuhira2002Dy2Ti2O7}, which does not undergo magnetic ordering. In this case, the plateau is due to the peculiar pyrochlore lattice structure with extreme Ising anisotropy at each of the 4 Dy sites having threefold symmetry long one of the $\langle 111 \rangle$ axes. 

In EuRhAl$_4$Si$_2$ and EuIrAl$_4$Si$_2$, the presence of such an unusual low field plateau is probably due to a combination of a rather low (but sizeable) crystalline anisotropy and of a peculiar ferrimagnetic\--like magnetic structure, which we believe consists in a tripled magnetic unit cell along $c$ with ``up-up-down'' moment arrangement. The propagation vetor would then be $k$=(0 0 1/3). Neutron diffraction experiments on a single crystal of EuRhAl$_4$Si$_2$ are planned in order to check the existence of such a structure. 
\begin{figure}[h]
\vspace{0.5cm}
\includegraphics[width=0.35\textwidth]{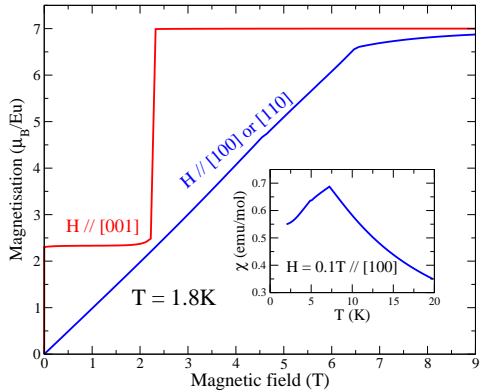}
\caption{\label{eurh42_mH}(Color online) Calculated isothermal magnetization curves at 1.8\,K for $H~\parallel$~[001] and in the $(ab)$ plane with $J_0$=0.18\,K, $J_1=-$0.02\,K, $J_2=-$0.17\,K and $D=-$1.5\,K (see text). Insert: calculated susceptibility for $H~\parallel$~[100] with the same parameters. These simulations are to be compared with the data in EuRhAl$_4$Si$_2$ in Figs.\ref{Chi_T_low_T}a and \ref{MH}a.}
\end{figure}
We present in the following a model which sets our assumption of an ``up-up-down'' sequence, with moments along $c$, on a firmer ground. The model considers 6 magnetic sub\--lattices consisting of ferromagnetic planes perpendicular to [001], and the Eu ions are taken to interact {\it via} three exchange integrals: an in\--plane nearest neighbour $J_0$, and interplane (along $c$) first neighbour $J_1$ and second neighbour $J_2$. We also introduce in the model the infinite range dipolar field on each ion due to its own sub\--lattice and to the 5 other sub\--lattices, using Ewald\--type summations, and a crystalline anisotropy term $DS_z^2$, with $D<0$ so that the [001] axis is an easy axis. As underlined in Refs.\onlinecite{Maurya2013EuNiGe3,rotter2003dipole}, inclusion of the dipolar interaction is essential to reproduce the magnetic behaviour in a realistic way in compounds with $S$-state Eu$^{2+}$ and Gd$^{2+}$ ions because it competes with the crystalline anisotropy for establishing the moment direction and the spin-flop fields. We perform a self\--consistent calculation of the magnetisation and of the susceptibility in the mean field approximation within the 6 sublattices \cite{Maurya2013EuNiGe3}. We find that the crystalline anisotropy term $D$ must be taken rather large (negative) in order to have moments along $c$, otherwise the dipolar field forces the moment to lie in the ($ab$) plane. By scanning the 4 parameter space \{$J_0$, $J_1$, $J_2$, $D$\}, we obtain for both compounds a point around which the zero field ground state is the ``up-up-down'' structure, and for which the magnetisation curves are in rather good agreement with experiment as to the low field jump, the spin-flop and spin-flip fields. However, the $T_{\rm N}$ values are somewhat lower (by 2-3\,K) than the actual values. For EuRhAl$_4$Si$_2$, these parameters are: $J_0$=0.18\,K, $J_1$=$-0.02$\,K, $J_2$=$-0.17$\,K and $D$=$-$1.3\,K. 
The corresponding $m(H)$ and $\chi(T)$ curves are shown in Fig.\ref{eurh42_mH}, where it can be seen that $T_N \simeq$7.5\,K, instead of the experimental value 10-11\,K. The nearest neighbour in-plane exchange integral is positive ferromagnetic, ensuring ferromagnetic planes normal to [001], and the farther neighbour exchange integrals are negative antiferromagnetic. Similarly, the parameter set: $J_0$=0.28\,K, $J_1$=$-$0.05\,K, $J_2$=$-$0.25\,K and $D$=$-$1.5\,K reproduces well the data in EuIrAl$_4$Si$_2$, but with $T_{\rm N}$~=~12\,K instead of 13-14\,K. For this latter compound, $J_0$ and $\vert J_2 \vert$ are larger than for EuRhAl$_4$Si$_2$, which accounts for the higher N\'eel temperature and characteristic field values. 

\section{Conclusion}

We have performed a series of measurements to probe in detail the magnetic properties of single crystals of two tetragonal quaternaries, EuRhAl$_4$Si$_2$ and EuIrAl$_4$Si$_2$. Both present a first order transition from a paramagnetic to an incommensurate antiferromagnetic phase at 11.7 and 14.7\,K respectively, followed by a lock\--in transition to a commensurate antiferromagnetic phase at 10.4 and 13.2\,K. Such a transition cascade is rather common among Eu intermetallics. By contrast, the magnetic behavior deep in the antiferromagnetic phase is rather unexpected for isotropic divalent Eu: the response is antiferromagnetic for a field applied in the $ab$\--plane and ferromagnetic for a field applied along $c$. In this latter case, the magnetization shows a low field jump followed by an intermediate plateau region attaining a value of 1/3 M$_0$. Beyond the plateau region there is a sharp hysteretic spin\--flip transition. Our in\--field transport data correlate well with this behaviour of the magnetisation. It hints to the presence of an unusual ``up-up-down'' zero field magnetic structure which is shown to be obtainable by the mean field model presented here, but which must be confirmed by neutron diffraction experiments.

\bibliography{ref}
\bibliographystyle{apsrev4-1}

\end{document}